\documentclass[english,floatfix,superscriptaddress,prb,twocolumn,showpacs]{revtex4}
\usepackage[T1]{fontenc}
\usepackage[latin9]{inputenc}
\usepackage{float}
\usepackage{amsmath}
\usepackage{graphicx}
\usepackage{amssymb}

\makeatletter
\usepackage{amsfonts}

\usepackage{amsfonts}

\makeatother

\usepackage{babel}

\begin{document}

\title{Energy-resolved spatial inhomogeneity of disordered Mott systems}

\author{E. C. Andrade}

\affiliation{Department of Physics and National High Magnetic Field Laboratory,
Florida State University, Tallahassee, FL 32306}

\affiliation{Instituto de F\i\'{}sica Gleb Wataghin, Unicamp, C.P. 6165, Campinas,
SP 13083-970, Brazil}

\author{E. Miranda}

\affiliation{Instituto de F\i\'{}sica Gleb Wataghin, Unicamp, C.P. 6165, Campinas,
SP 13083-970, Brazil}

\author{V. Dobrosavljevi\'{c}}

\affiliation{Department of Physics and National High Magnetic Field Laboratory,
Florida State University, Tallahassee, FL 32306}

\date{\today}

\begin{abstract}
We investigate the effects of weak to moderate disorder on the $T=0$
Mott metal-insulator transition in two dimensions. Our model calculations
demonstrate that the electronic states close to the Fermi energy become
more spatially homogeneous in the critical region. Remarkably, the
higher energy states show the opposite behavior: they display enhanced
spatial inhomogeneity precisely in the close vicinity to the Mott
transition. We suggest that such energy-resolved disorder screening
is a generic property of disordered Mott systems.
\end{abstract}


\maketitle

\section{Introduction}

The observation of a metal-insulator transition in high-mobility two-dimensional
electron systems at zero magnetic field has sparked renewed interest
in this type of transition \cite{rmp_2d_mit}. In these systems, electron-electron
interactions represent the largest energy scale in the problem \cite{rmp_2d_mit}.
Further evidence for the crucial role of electronic correlations has
come from complementary experiments \cite{pudalov_2d_mit,kravchenko_2d_mit}
reporting a substantial mass enhancement close to the metal-insulator
transition. Taken together these experimental results stress the importance
of the ill-understood effects of disorder in strongly correlated electronic
systems \cite{lee_ramakrishnan}.

In one of the first studies of its kind, Tanaskovi\'{c} \emph{et
al.} have investigated the interplay of strong correlations and disorder
using a dynamical mean field theory (DMFT) \cite{dmft_rmp96} approach
\cite{Screening_2003}. The DMFT approach to disordered systems treats
correlations on a local level by solving the embedded-atom strongly
correlated Anderson impurity problem in the self-consistently determined
\emph{fixed} bath of the other electrons. One is thus forced to consider
an \emph{ensemble} of single-impurity actions, one for each lattice
site. Remarkably, Tanaskovi\'{c} \emph{et al.} found that very strong
site disorder screening emerges precisely in the vicinity of the Mott
metal-insulator transition \cite{mott_mit}. This effect can be traced
back to the pinning of the single-impurity Kondo resonances to the
Fermi level, which acts to suppress the effective randomness.

Motivated by this striking result, we extended their work to finite
dimensions\emph{.} To treat both strong correlations and disorder
in a non-perturbative fashion at $T=0$, we use a generalization of
the DMFT, the statistical DMFT (\textit{\emph{statDMFT}}) \cite{statDMFT},
implemented using a slave boson impurity solver \cite{KRSB4}. This
approach is equivalent to the description of the effects of disorder
through a Gutzwiller-type wave function \cite{gutzwiller63,gutzwiller65},
which in the clean case realizes the Brinkman-Rice scenario of the
Mott transition \cite{Brinkman_Rice}. The statDMFT method retains
the local treatment of electronic correlations. However, in contrast
to the infinite-dimensional DMFT approach, here each strongly correlated
site sees a \emph{different bath of electrons}, reflecting the strong
spatial fluctuations of their immediate environment. We find that,
concomitant to a strong pinning effect, additional effective disorder
is also generated through an increasingly broader distribution of
quasiparticle weights. Their interplay leads to a non-trivial landscape
in energy space: the proximity to the Mott transition acts to suppress
density of states fluctuations close to the Fermi level, while at
the same time \emph{enhancing} them at higher energies.

\section{Strongly correlated theory}

We focus on the disordered Hubbard model \begin{eqnarray}
H & = & \sum_{i,\sigma}\varepsilon_{i}c_{i\sigma}^{\dagger}c_{i\sigma}-t\sum_{\left\langle ij\right\rangle ,\sigma}\left(c_{i\sigma}^{\dagger}c_{j\sigma}+\mathrm{h.c.}\right)\nonumber \\
 & + & U\sum_{i}n_{i\uparrow}n_{i\downarrow},\label{eq:Hubbard_disordered}\end{eqnarray}
where $c_{i\sigma}^{\dagger}$$\left(c_{i\sigma}\right)$ is the creation
(annihilation) operator of an electron with spin projection $\sigma$
on site $i$, $-t$ is the nearest-neighbor hopping amplitude, $U$
is the on-site Hubbard repulsion, $n_{i\sigma}=c_{i\sigma}^{\dagger}c_{i\sigma}$
is the number operator, and the site energies $\varepsilon_{i}$ are
uniformly distributed in the interval $\left[-W/2\mbox{,}\, W/2\right]$,
where $W$ is the disorder strength. We work at half filling (chemical
potential $\mu=U/2$) on an $L\mbox{X}L$ square lattice with periodic
boundary conditions. All energies will be expressed in units of the
clean Fermi energy (the half-bandwidth $D$) $E_{F}=4t$.

We treat the Hamiltonian of Eq.~\ref{eq:Hubbard_disordered} in its
paramagnetic phase within the statDMFT \cite{statDMFT}. This theory
is exact in the non-interacting limit and reduces to the standard
DMFT in the absence of disorder. Unlike the DMFT, however, it incorporates
Anderson localization effects. We start by writing an effective action
(in imaginary time) for a given site $i$, with the simplification
that we neglect all non-quadratic terms in the local fermionic operators
except for the local $U$-term\begin{eqnarray}
S_{eff}^{\left(i\right)} & = & \sum_{\sigma}\int_{0}^{\beta}d\tau c_{i\sigma}^{\dagger}\left(\tau\right)\left(\partial_{\tau}+\varepsilon_{i}-\mu\right)c_{i\sigma}\left(\tau\right)\nonumber \\
 & + & \sum_{\sigma}\int_{0}^{\beta}d\tau\int_{0}^{\beta}d\tau^{\prime}c_{i\sigma}^{\dagger}\left(\tau\right)\Delta_{i}\left(\tau-\tau^{\prime}\right)c_{i\sigma}\left(\tau^{\prime}\right)\nonumber \\
 & + & U\int_{0}^{\beta}d\tau n_{i\uparrow}\left(\tau\right)n_{i\downarrow}\left(\tau\right)\mbox{.}\label{eq:Action_Stat_DMFT}\end{eqnarray}
The site $i$ is connected with the rest of the lattice through the
bath (or ``cavity'') function $\Delta_{i}\left(\tau\right)$, which
in statDMFT (but in contrast to DMFT) varies from site to site and
thus exhibits strong spatial fluctuations. 

The effective action in Eq.~\ref{eq:Action_Stat_DMFT} is precisely
the action of an Anderson impurity model \cite{aim_1961} embedded
in a sea of conduction electrons described by $\Delta_{i}\left(\tau\right)$.
Therefore, this approach maps the original Hubbard Hamiltonian in
Eq.~\ref{eq:Hubbard_disordered} onto an \emph{ensemble} of single-impurity
Anderson Hamiltonians \cite{georges_kotliar_aim}. The local $i$-site
Green's function, calculated under the dynamics dictated by the effective
action in Eq.~\ref{eq:Action_Stat_DMFT}, can be written as ($i\omega$
is a Matsubara frequency)
\begin{equation}
G_{i}^{loc}\left(i\omega\right)  =  \frac{1}{i\omega+\mu-\varepsilon_{i}-\Sigma_{i}\left(i\omega\right)-\Delta_{i}\left(i\omega\right)},\label{eq:G_loc_imp}
\end{equation}
which also serves as a definition of $\Sigma_{i}\left(i\omega\right)$,
the $i$-site self-energy. It is important to point out that within
statDMFT the electronic self-energy $\Sigma_{i}\left(i\omega\right)$
is still local, albeit site-dependent. 

The bath function $\Delta_{i}\left(i\omega\right)$ can be viewed
as the Weiss field of this mean field theory, here elevated to a full
function of frequency or time. It is determined through a self-consistency
condition that demands that the Green's function $G_{i}^{loc}\left(i\omega\right)$
obtained from the effective action in Eq.~\ref{eq:Action_Stat_DMFT}
be equal to the diagonal (local) part of the full lattice Green's
function 
\begin{equation}
G_{ii}\left(i\omega\right)  =  \left[\frac{1}{i\omega-\mbox{\boldmath$\varepsilon$}-\mathbf{H}_{0}-\mbox{\boldmath$\Sigma$}\left(i\omega\right)}\right]_{ii},\label{eq:G_lattice}
\end{equation}
where $\mbox{\boldmath$\Sigma$}\left(i\omega\right)$ and $\mbox{\boldmath$\varepsilon$}$
are site-diagonal matrices $\left[\mbox{\boldmath$\Sigma$}\left(i\omega\right)\right]_{ij}=\Sigma_{i}\left(i\omega\right)\delta_{ij}$,
$\left[\mbox{\boldmath$\varepsilon$}\right]_{ij}=\varepsilon_{i}\delta_{ij}$
and $\mathbf{H}_{0}$ is the clean ($W=0$) and non-interacting ($U=0$)
lattice Hamiltonian. In general, this step involves the inversion
of the frequency-dependent matrix within brackets in Eq.~(\ref{eq:G_lattice}).

It is worthwhile to point out that the statDMFT approach requires
a massive numerical effort since it is necessary to solve a single
impurity problem \emph{for every lattice site} as well to perform
the inversion implied by the self-consistency condition. On the other
hand, it provides access to entire distribution functions and accounts
for spatial correlations between local quantities. 

To find $\Sigma_{i}\left(i\omega\right)$, we need to solve the auxiliary
single impurity problems for a given set of $\Delta_{i}\left(i\omega\right)$.
For this task, we have used the four-boson mean-field theory of Kotliar
and Ruckenstein \cite{KRSB4} at $T=0$, which is equivalent to the
well-known Gutzwiller variational approximation. In practice, we need
to solve a pair of non-linear equations for the site-dependent Kotliar-Ruckenstein
slave boson amplitudes $e_{i}$ and $d_{i}$ \cite{KRSB4,Screening_2003}
\begin{eqnarray}
2\int_{-\infty}^{\infty}\frac{d\omega}{2\pi}G_{i}^{loc}\left(i\omega\right) & = & Z_{i}\left(1-e_{i}^{2}+d_{i}^{2}\right),\label{eq:SB1}\\
\int_{-\infty}^{\infty}\frac{d\omega}{2\pi}\Delta_{i}\left(i\omega\right)G_{i}^{loc}\left(i\omega\right) & = & \frac{Z_{i}e_{i}\left(\varepsilon_{i}-Z_{i}v_{i}-\mu\right)}{\partial Z_{i}/\partial e_{i}},\label{eq:SB2}\end{eqnarray}
where
\begin{equation}
Z_{i}  =  \frac{2\left(e_{i}+d_{i}\right)^{2}\left(1-e_{i}^{2}-d_{i}^{2}\right)}{1-\left(e_{i}^{2}-d_{i}^{2}\right)^{2}},\label{eq:Z_i}\end{equation}
and \begin{equation}
Z_{i}v_{i}  =  \varepsilon_{i}-\mu+\frac{Ud_{i}}{\left(\frac{\partial Z_{i}}{\partial e_{i}}d_{i}+\frac{\partial Z_{i}}{\partial d_{i}}e_{i}\right)}\frac{\partial Z_{i}}{\partial e_{i}}.\label{eq:v_i}\end{equation}
Eqs.~(\ref{eq:SB1}-\ref{eq:SB2}) involve the local $i$-site Green's
function $G_{i}^{loc}\left(i\omega\right)$. In the Kotliar-Ruckenstein
theory, the $i$-site self-energy $\Sigma_{i}\left(i\omega\right)$
is given by
\begin{equation}
\Sigma_{i}\left(i\omega\right)  =  \left(1-Z_{i}^{-1}\right)i\omega+v_{i}-\varepsilon_{i}+\mu,\label{eq:self_energy_SB4}\end{equation}
which, when plugged into Eq.~(\ref{eq:G_loc_imp}) yields\begin{equation}
G_{i}^{loc}\left(i\omega\right)  =  \frac{Z_{i}}{i\omega-Z_{i}v_{i}-Z_{i}\Delta_{i}\left(i\omega\right)}.\label{eq:G_loc_impKR}\end{equation}
It is clear that $Z_{i}$ has the physical interpretation of a quasi-particle
weight (wave function renormalization). It also renormalizes the hybridization
function, thus setting the local Kondo temperature \cite{readnewns83,readnewns83b,colemanlong87}.
Furthermore, from Eq.~(\ref{eq:self_energy_SB4}) we see that $v_{i}$
can be viewed as a renormalized on-site disorder potential
\begin{equation}
v_{i}  =\varepsilon_{i}+\Sigma_{i}\left(0\right)-\mu.\label{eq:ren_dis_pot}
\end{equation}

We numerically solved the disordered Hubbard model with statDMFT,
using the Kotliar-Ruckenstein theory as the impurity solver, for several
lattice sizes up to $L=50$. For every $\left(U,W\right)$ pair we
typically generated around forty realizations of disorder. We carefully
verified that for such large lattices, all our results are robust
and essentially independent of the system size (see, e.g., the inset
of Fig.~\ref{fig:P(v)}).

\begin{figure}[t]
\bigskip{}

\begin{centering}\includegraphics[bb=53bp 40bp 714bp 530bp,scale=0.25]{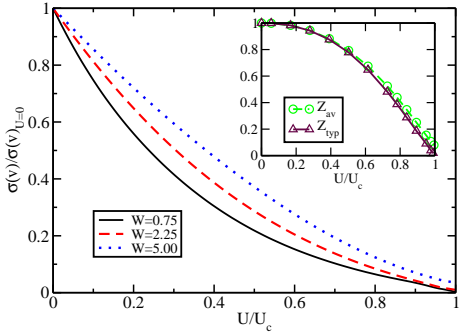} \par\end{centering}

\caption{\label{fig:Screening}Strength of the renormalized site energy disorder,
as given by the standard deviation $\sigma\left(v\right)$ of $P\left(v\right)$,
normalized by its non-interacting value (see text for definition of
$v$). Close to the Mott transition, the disorder screening remains
strong even for moderate values of disorder. Results are shown for
$L=20$. In the inset we show the typical ($Z_{typ}$) and average
($Z_{av}$) values of the local quasiparticle weight $Z_{i}$ as functions
of the interaction $U$. The Mott transition is identified by the
vanishing of $Z_{typ}$. We note that $Z_{av}$ is finite at $U_{c}$
indicating that a fraction of the sites remains nearly empty or doubly
occupied. Results are shown for $L=20$ and $W=2.25$.}
\end{figure}

\section{The disordered Mott transition}

According to the scaling theory of localization, any amount of disorder
drives a system with dimension equal to or smaller than two to an
insulating phase \cite{gang4}. However, this result was obtained
in the absence of electron-electron interactions. ln the last few
years, strong numerical evidence has been obtained indicating that
interactions can act to enhance the conducting properties of two-dimensional
electronic systems \cite{nandini_prl2004,scalettar_prb2007,statDMFT_rwortis}.
Moreover, the stability of a $2d$ metal with respect to weak-localization
corrections has been investigated by Punnoose and Finkelstein \cite{punnoose_sci2005}
in very careful recent work. These authors have demonstrated that
any $2d$ metal remains stable with respect to sufficiently weak disorder
due to additional (anti-localizing) interaction corrections. Note,
however, that both the weak-localization and the corresponding interaction
corrections are manifested only through a very weak, logarithmic dependence
on the system size. Such subtle finite size effects are not visible
for the very weak (renormalized) disorder we deal with in this work.
Nevertheless, based on the very convincing considerations of Punnoose
and Finkelstein, the stability of a $2d$ metal is beyond immediate
doubt and these issues are not of relevance for the questions we focus
on in this paper.

The low energy behavior on the metallic side of the transition is
characterized by the distribution of the local quasiparticle weight
$Z_{i}$. Following previous studies of the disordered Mott transition
\cite{statDMFT,Screening_2003,imp_scaling_prb_2006}, we choose the
typical value of $Z_{i}$, here defined through the geometrical average
$Z_{typ}=\mbox{exp}\left\{ \left\langle \mbox{ln}Z_{i}\right\rangle \right\} $,
as the order parameter of the transition. This quantity vanishes linearly
at a critical value of interaction $U_{c}\equiv U_{c}\left(W\right)$
at which the Mott transition takes place, marking the transmutation
of most itinerant electrons into local magnetic moments (see the inset
of Fig.~\ref{fig:Screening}). Because random site energies tend
to push the local occupation away from half filling, $U_{c}(W)$ is
an increasing function of the disorder strength $W$ \cite{griffiths_2d}.
We also stress that the average value of the quasiparticle weight
$Z_{av}$ is small yet finite at the Mott transition (inset of Fig.~\ref{fig:Screening}),
indicating that some sites remain either empty ($e_{i}=1$) ou doubly
occupied ($d_{i}=1$) and do not give rise to localized magnetic moments.

\begin{figure}[b]
\bigskip{}

\begin{centering}\includegraphics[bb=53bp 40bp 714bp 530bp,scale=0.25]{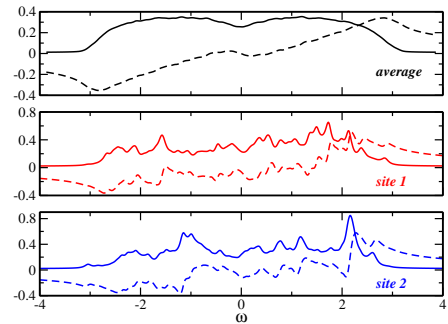} \par\end{centering}

\caption{\label{fig:hibfxw}Real (dashed lines) and imaginary (full lines)
parts of the hybridization function $\Delta_{i}\left(\omega\right)$.
Although the average value of $\Delta_{i}\left(\omega\right)$ is
particle-hole symmetric, for an arbitrary site $\Delta_{i}\left(\omega\right)$
is locally particle-hole asymmetric. Results are shown for $L=20$
and $W=U=2.25$.}
\end{figure}

\section{Particle-hole symmetry and strong disorder screening}

Using the fact that in the current statDMFT approach the lattice problem
is mapped onto an \emph{ensemble} of auxiliary Anderson impurity problems,
we can characterize the approach to the Mott transition by a steady
decrease of the local Kondo temperature $T_{K}^{i}$ $\left(T_{K}^{i}\propto Z_{i}\right)$.
Moreover, the renormalized disorder potential $v_{i}$ can be thought
of as the position of the local Kondo resonance energy.

In the DMFT limit $\left(d\rightarrow\infty\right)$, each site has
many neighbors and thus is embedded in the same (self-averaged) environment
described by $\Delta_{av}\left(\omega\right)$. In this regime, as
long as $\Delta_{av}\left(\omega\right)$ is particle-hole symmetric,
we find \emph{perfect} disorder screening close to the Mott transition
\cite{Screening_2003}. This happens because, as we increase $U$
towards $U_{c}\left(W\right)$, we approach the Kondo limit, $Z_{i}\rightarrow0$,
causing $v_{i}$ (the Kondo resonance) to be ``pinned'' to the Fermi
energy \cite{nozieres74}.

\begin{figure}[t]
\bigskip{}

\begin{centering}\includegraphics[bb=53bp 40bp 714bp 530bp,scale=0.25]{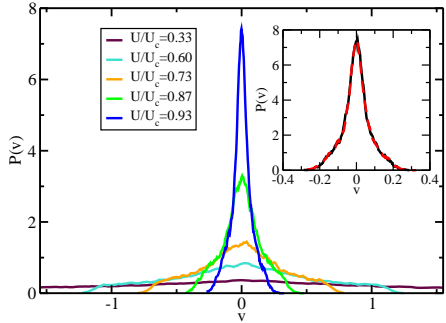} \par\end{centering}

\caption{\label{fig:P(v)}Probability distribution function $P\left(v\right)$
of the renormalized site energy $v$. As we move towards the Mott
transition, disorder screening takes place and $P\left(v\right)$
becomes increasingly narrower. Results are shown for $L=50$ and $W=2.25$.
The inset illustrates how for such large lattices our results for
$P\left(v\right)$ are essentially independent of the system size:
results are shown for $L=20$, full line, and $L=50$, dashed line
at $W=2.25$ and $U/U_{c}=0.93$. }
\end{figure}

However, within the \textit{\emph{statDMFT}} approach, $\Delta_{i}\left(\omega\right)$
fluctuates strongly from site to site and is not locally particle-hole
symmetric (see Fig.~\ref{fig:hibfxw}). For this reason, we expect
that $v_{i}$ will also have a contribution which is proportional
to $\mbox{Re}\left[\Delta_{i}\left(0\right)\right]$. Therefore, we
have no guarantee that a similar mechanism of disorder screening will
persist close to the critical point in $2d$. Surprisingly, however,
for small and moderate disorder $\left(W\lesssim U_{c}\right)$, we do
get a very strong disorder screening close to the Mott metal-insulator
transition (see Figs.~\ref{fig:Screening} and \ref{fig:P(v)}).

\begin{figure}[b]
\bigskip{}

\begin{raggedright}\includegraphics[bb=53bp 40bp 714bp 530bp,scale=0.4]{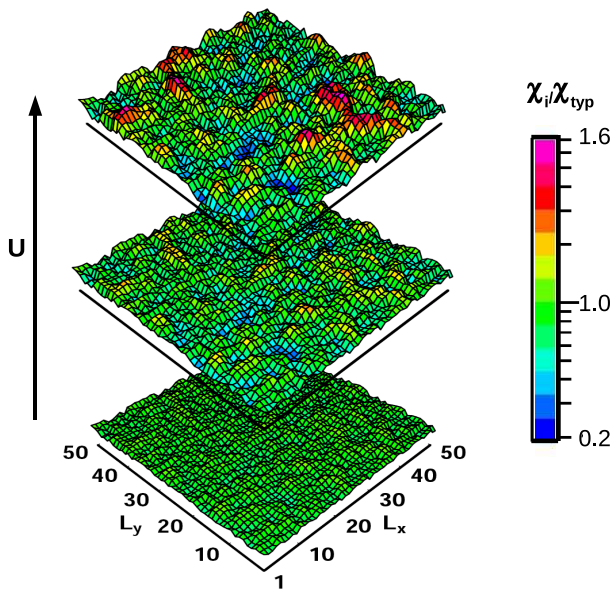} \par\end{raggedright}

\caption{\label{fig:ZZ_spatial}Spatial distribution of the of the local susceptibility
$\chi_{i}\sim Z_{i}^{-1}$ normalized by its typical value $\chi_{typ}$.
From bottom to the top we have $U/U_{c}=0.73,\,0.87,\,0.96$. The
color scale is logarithmic in order to stress that, as we approach
the critical point, we have the formation of regions in which $\chi_{i}\gg\chi_{typ}$
(localized magnetic moments) and $\chi_{i}\ll\chi_{typ}$ (Anderson
insulator droplets). Results are shown for $W=2.25$ and $L=50$. }
\end{figure}

\section{Mott droplets}

The site-to-site fluctuations in the hybridization function encode
\emph{spatial correlations} between the physical quantities, implying,
for example, that the value of the local quasi-particle weight at
a given site depends on the quasi-particle weight values at the neighboring
sites. Therefore, due to rare disorder configurations, it is possible
to find regions containing sites in which $Z_{i}\ll Z_{typ}$ (red
regions in Fig.~\ref{fig:ZZ_spatial}). Since the approach to the
Mott insulator corresponds to $Z\rightarrow0$, such regions with
$Z_{i}\ll Z_{typ\text{ }}$ should be recognized as {}``almost localized''
Mott droplets (consisting of localized magnetic moments \cite{milovanovic_prl1989,paalanen_prl1988,vlad_kotliar_prl1993})
inside the strongly correlated metallic host. Within our Brinkman-Rice
picture, each local region provides \cite{statDMFT,Brinkman_Rice}
a contribution $\chi_{i}\sim\gamma_{i}\sim Z_{i}^{-1}$ to the spin
susceptibility or the Sommerfeld coefficient, respectively. The local
regions with the smallest $Z_{i}$ thus dominate the thermodynamic
response and will ultimately give rise to an Electronic Griffiths
Phase in the vicinity of the disordered Mott metal-insulator transition
\cite{griffiths_2d,NFL_2005,tvojta_jpa}. 

\begin{figure}[t]
\bigskip{}

\begin{centering}\includegraphics[bb=53bp 40bp 714bp 530bp,scale=0.25]{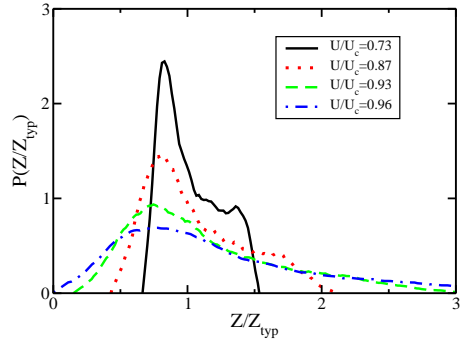} \par\end{centering}

\caption{\label{fig:P(Z_typ)}Distribution of the quasiparticle weight divided
by its typical value for $W=2.25$ and $U_{c}\left(W\right)\simeq3.7$.
As we approach the critical point, $P\left(Z/Z_{typ}\right)$ becomes
increasingly broader, even though the bare disorder strength $W$
is kept fixed. Results are shown for $L=20$. }
\end{figure}

The Mott droplets have a direct influence on the probability distribution
function $P\left(Z/Z_{typ}\right)$, since they give rise to a low-$Z$
tail, contributing to the fact that $P\left(Z/Z_{typ}\right)$ actually
broadens as we approach the Mott metal-insulator transition for a
\emph{fixed} $W$, as shown in Figs. \ref{fig:ZZ_spatial} and \ref{fig:P(Z_typ)}. 

Surely, there is also a contribution to the broadening of $P\left(Z/Z_{typ}\right)$
coming from the high-$Z$ tail originating from those sites with $Z_{i}\gg Z_{typ}$.
Such sites retain a finite $Z_{i}$ at the transition and are nearly
empty or doubly occupied, giving rise to Anderson insulating regions
(blue regions in Fig.~\ref{fig:ZZ_spatial}). The coexistence of
localized magnetic moments, and nearly empty or doubly occupied sites
close to the critical point is characteristic of a two-fluid behavior \cite{imp_scaling_prb_2006}. Nevertheless,
this interesting phenomenon is not relevant for the present analysis,
since the major contribution to the energy resolved inhomogeneities,
discussed in the next Section, as well as to the thermodynamic response,
comes from the Mott droplets. 

\section{Energy-resolved inhomogeneities} 

From the above discussion, it is clear that the behaviors of $P\left(v\right)$
and $P\left(Z/Z_{typ}\right)$ near the Mott metal-insulator transition
are quite distinct. While the former exhibits strong disorder screening
(Fig.~\ref{fig:P(v)}), the latter suggests that the disorder is
actually increasing (Fig.~\ref{fig:P(Z_typ)}). This dichotomy gives
rise to an \emph{energy-dependent effective disorder}, which manifests
itself, for example, in the spatial structure of the local density
of states. The local density of states is defined as $\rho_{i}\left(\omega\right)=\left(1/\pi\right)\mbox{Im}\left[G_{ii}\left(\omega-i0^{+}\right)\right]$,
where the lattice Green's function was given in Eq.~\ref{eq:G_lattice},
and has the following expression within our Brinkman-Rice picture
\begin{equation}
G_{ii}\left(\omega\right)=\left[\frac{1}{\mathbf{Z}^{-1}\omega-\mathbf{v}-\mathbf{H_{0}}}\right]_{ii},\label{eq:G_lattice_SB4}\end{equation}
where $\mathbf{Z}$ and $\mathbf{v}$ are site-diagonal matrices such
that $\left[\mathbf{Z}\right]_{ij}=Z_{i}\delta_{ij},\,\left[\mathbf{v}\right]_{ij}=v_{i}\delta_{ij}$.
Therefore, the frequency-dependent effective disorder potential ``seen''
by the quasiparticles at energy $\omega$ can be defined as \begin{equation}
\varepsilon_{i}^{eff}(\omega)=v_{i}-\frac{\omega}{Z_{i}}.\label{eq:effec_dis_pot}\end{equation}

\begin{figure}[h]
\bigskip{}

\begin{centering}\includegraphics[scale=0.35]{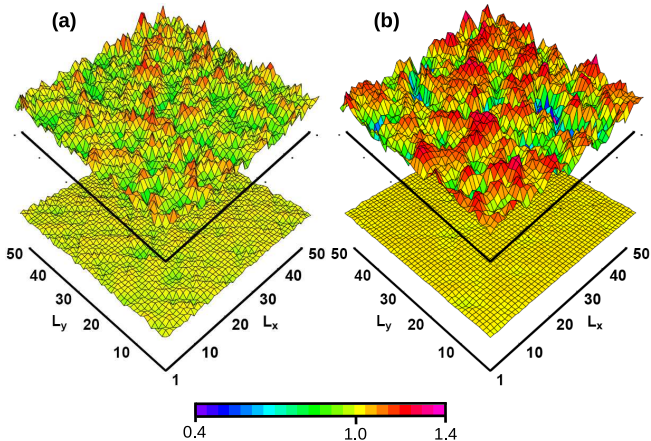} \par\end{centering}

\caption{\label{fig:ldos_spatial}Spatial distribution of the local density
of states normalized by its clean and non-interacting value for one
given realization of disorder and two distinct values of interaction:
\emph{(a)} $U/U_{c}=0.87$ and \emph{(b)} $U/U_{c}=0.96$, where $U_{c}\simeq3.7$.
We have $\omega=0$ in the bottom figures and $\omega/Z_{typ}=0.1$
in the top ones. At the Fermi energy, the local density of states
distribution becomes homogeneous as we approach the Mott transition.
Conversely, if we move even slightly away from the Fermi energy, the
distribution becomes in fact more inhomogeneous close to $U_{c}$.
Results are shown for $L=50$, $W=2.25$. }
\end{figure}

In our Brinkman-Rice scenario of the Mott transition, the quasi-particle
bandwidth is reduced by $Z_{typ}$ as $U\rightarrow U_{c}$. To monitor
the behavior within the quasi-particle band, we therefore introduce
a rescaled frequency $\omega^{\ast}=\omega/Z_{typ}$, which we will
keep constant as we approach the transition. In Fig.~\ref{fig:ldos_spatial},
we show topographic maps of the local density of states for one specific
realization of disorder. Because of strong disorder screening, $v_{i}\approx0$
close to the critical point. Thus, if the system is examined at the
Fermi energy ($\omega=0$), it becomes more and more homogeneous as
the transition is approached. At higher energies ($\omega^{\ast}\neq0$),
however, the fluctuations in $Z_{typ}/Z_{i}$ come into play. Since
they are very pronounced close to the Mott transition, we instead
find a strong enhancement of the spatial inhomogeneity. 

This result is surprisingly reminiscent of recent spectroscopic images
on doped cuprates \cite{seamus_davis_sci2005}. Our theory, which
focuses on local (Kondo-like) effects of strong correlations (while
neglecting inter-site magnetic correlations) and does not include
any physics associated with superconducting pairing, strongly suggests
that such energy-resolved inhomogeneity is a robust and general feature
of disordered Mott systems. Indeed, recent results corroborate this
picture of an energy-resolved strong-correlation driven disorder screening
\cite{gargetal08}.

\section{Conclusions}

We have discussed the results of a Brinkman-Rice approach to the disordered
Mott transition. A striking feature that is apparent in this scenario
is the different behaviors of the effective site disorder and the
quasiparticle weights as the transition is approached. Whereas randomness
in the former is suppressed, the latter becomes extremely singular
and broad on the way to the Mott insulator. The end result of this
dichotomy is a non-trivial energy-space landscape, which is signaled
by an energy-dependent inhomogeneity.

\section{Acknowledgments}

This work was supported by FAPESP through grants 04/12098-6 (ECA) and
07/57630-5 (EM), CAPES through grant 1455/07-9 (ECA), CNPq through
grant 305227/2007-6 (EM), and by NSF through grant DMR-0542026 (VD).

%\bibliographystyle{apsrev}
%\bibliography{proceeding}

\begin{thebibliography}{90}
\expandafter\ifx\csname natexlab\endcsname\relax\def\natexlab#1{#1}\fi
\expandafter\ifx\csname bibnamefont\endcsname\relax
  \def\bibnamefont#1{#1}\fi
\expandafter\ifx\csname bibfnamefont\endcsname\relax
  \def\bibfnamefont#1{#1}\fi
\expandafter\ifx\csname citenamefont\endcsname\relax
  \def\citenamefont#1{#1}\fi
\expandafter\ifx\csname url\endcsname\relax
  \def\url#1{\texttt{#1}}\fi
\expandafter\ifx\csname urlprefix\endcsname\relax\def\urlprefix{URL }\fi
\providecommand{\bibinfo}[2]{#2}
\providecommand{\eprint}[2][]{\url{#2}}

\bibitem{rmp_2d_mit}
\bibinfo{author}{\bibfnamefont{E.}~\bibnamefont{Abrahams}},
  \bibinfo{author}{\bibfnamefont{S.~V.} \bibnamefont{Kravchenko}},
  \bibnamefont{and} \bibinfo{author}{\bibfnamefont{M.~P.}
  \bibnamefont{Sarachik}}, \bibinfo{journal}{Rev.\ Mod.\ Phys.}
  \textbf{\bibinfo{volume}{73}}, \bibinfo{pages}{251} (\bibinfo{year}{2001}).

\bibitem{pudalov_2d_mit}
\bibinfo{author}{\bibnamefont{{{V. M. Pudalov} {\it et al.}}}},
  \bibinfo{journal}{Phys.\ Rev.\ Lett.} \textbf{\bibinfo{volume}{88}},
  \bibinfo{pages}{196404} (\bibinfo{year}{2002}).

\bibitem{kravchenko_2d_mit}
\bibinfo{author}{\bibnamefont{{{A. A. Shashkin} {\it et al.}}}},
  \bibinfo{journal}{Phys.\ Rev.\ B} \textbf{\bibinfo{volume}{66}},
  \bibinfo{pages}{073303} (\bibinfo{year}{2002}).

\bibitem{lee_ramakrishnan}
\bibinfo{author}{\bibfnamefont{P.~A.} \bibnamefont{Lee}} \bibnamefont{and}
  \bibinfo{author}{\bibfnamefont{T.~V.} \bibnamefont{Ramakrishnan}},
  \bibinfo{journal}{Rev.\ Mod.\ Phys.} \textbf{\bibinfo{volume}{57}},
  \bibinfo{pages}{287} (\bibinfo{year}{1985}).

\bibitem{dmft_rmp96}
\bibinfo{author}{\bibfnamefont{A.}~\bibnamefont{Georges}},
  \bibinfo{author}{\bibfnamefont{G.}~\bibnamefont{Kotliar}},
  \bibinfo{author}{\bibfnamefont{W.}~\bibnamefont{Krauth}}, \bibnamefont{and}
  \bibinfo{author}{\bibfnamefont{M.}~\bibnamefont{Rozenberg}},
  \bibinfo{journal}{Rev.\ Mod.\ Phys.} \textbf{\bibinfo{volume}{68}},
  \bibinfo{pages}{13} (\bibinfo{year}{1996}).

\bibitem{Screening_2003}
\bibinfo{author}{\bibfnamefont{D.}~\bibnamefont{Tanaskovi{\'c}}},
  \bibinfo{author}{\bibfnamefont{V.}~\bibnamefont{Dobrosavljevi{\'c}}},
  \bibinfo{author}{\bibfnamefont{E.}~\bibnamefont{Abrahams}}, \bibnamefont{and}
  \bibinfo{author}{\bibfnamefont{G.}~\bibnamefont{Kotliar}},
  \bibinfo{journal}{Phys.\ Rev.\ Lett.} \textbf{\bibinfo{volume}{91}},
  \bibinfo{pages}{066603} (\bibinfo{year}{2003}).

\bibitem{mott_mit}
\bibinfo{author}{\bibfnamefont{N.~F.} \bibnamefont{Mott}},
  \emph{\bibinfo{title}{Metal-Insulator Transition}}
  (\bibinfo{publisher}{Taylor and Francis}, \bibinfo{address}{London},
  \bibinfo{year}{1990}).

\bibitem{statDMFT}
\bibinfo{author}{\bibfnamefont{V.}~\bibnamefont{Dobrosavljevi{\'c}}}
  \bibnamefont{and} \bibinfo{author}{\bibfnamefont{G.}~\bibnamefont{Kotliar}},
  \bibinfo{journal}{Phys.\ Rev.\ Lett.} \textbf{\bibinfo{volume}{78}},
  \bibinfo{pages}{3943} (\bibinfo{year}{1997}).

\bibitem{KRSB4}
\bibinfo{author}{\bibfnamefont{G.}~\bibnamefont{Kotliar}} \bibnamefont{and}
  \bibinfo{author}{\bibfnamefont{A.}~\bibnamefont{Ruckenstein}},
  \bibinfo{journal}{Phys.\ Rev.\ Lett.} \textbf{\bibinfo{volume}{57}},
  \bibinfo{pages}{1362} (\bibinfo{year}{1986}).

\bibitem{gutzwiller63}
\bibinfo{author}{\bibfnamefont{M.~C.} \bibnamefont{Gutzwiller}},
  \bibinfo{journal}{Phys. Rev. Lett.} \textbf{\bibinfo{volume}{10}},
  \bibinfo{pages}{159} (\bibinfo{year}{1963}).

\bibitem{gutzwiller65}
\bibinfo{author}{\bibfnamefont{M.~C.} \bibnamefont{Gutzwiller}},
  \bibinfo{journal}{Phys. Rev.} \textbf{\bibinfo{volume}{137}},
  \bibinfo{pages}{A1726} (\bibinfo{year}{1965}).

\bibitem{Brinkman_Rice}
\bibinfo{author}{\bibfnamefont{W.~F.} \bibnamefont{Brinkman}} \bibnamefont{and}
  \bibinfo{author}{\bibfnamefont{T.~M.} \bibnamefont{Rice}},
  \bibinfo{journal}{Phys.\ Rev.\ B} \textbf{\bibinfo{volume}{02}},
  \bibinfo{pages}{4302} (\bibinfo{year}{1970}).

\bibitem{aim_1961}
\bibinfo{author}{\bibfnamefont{P.~W.} \bibnamefont{Anderson}},
  \bibinfo{journal}{Phys.\ Rev.} \textbf{\bibinfo{volume}{124}},
  \bibinfo{pages}{41} (\bibinfo{year}{1961}).

\bibitem{georges_kotliar_aim}
\bibinfo{author}{\bibfnamefont{A.}~\bibnamefont{Georges}} \bibnamefont{and}
  \bibinfo{author}{\bibfnamefont{G.}~\bibnamefont{Kotliar}},
  \bibinfo{journal}{Phys.\ Rev.\ B} \textbf{\bibinfo{volume}{45}},
  \bibinfo{pages}{6479} (\bibinfo{year}{1992}).

\bibitem{readnewns83}
\bibinfo{author}{\bibfnamefont{N.}~\bibnamefont{Read}} \bibnamefont{and}
  \bibinfo{author}{\bibfnamefont{D.~M.} \bibnamefont{Newns}},
  \bibinfo{journal}{J. Phys. C} \textbf{\bibinfo{volume}{16}},
  \bibinfo{pages}{3273} (\bibinfo{year}{1983}{\natexlab{a}}).

\bibitem{readnewns83b}
\bibinfo{author}{\bibfnamefont{N.}~\bibnamefont{Read}} \bibnamefont{and}
  \bibinfo{author}{\bibfnamefont{D.~M.} \bibnamefont{Newns}},
  \bibinfo{journal}{J. Phys. C} \textbf{\bibinfo{volume}{16}},
  \bibinfo{pages}{L1055} (\bibinfo{year}{1983}{\natexlab{b}}).

\bibitem{colemanlong87}
\bibinfo{author}{\bibfnamefont{P.}~\bibnamefont{Coleman}},
  \bibinfo{journal}{Phys. Rev. B} \textbf{\bibinfo{volume}{35}},
  \bibinfo{pages}{5072} (\bibinfo{year}{1987}).

\bibitem{gang4}
\bibinfo{author}{\bibfnamefont{E.}~\bibnamefont{Abrahams}},
  \bibinfo{author}{\bibfnamefont{P.~W.} \bibnamefont{Anderson}},
  \bibinfo{author}{\bibfnamefont{D.~C.} \bibnamefont{Licciardelo}},
  \bibnamefont{and}
  \bibinfo{author}{\bibfnamefont{T.}~\bibnamefont{Ramakrishnan}},
  \bibinfo{journal}{Phys.\ Rev.\ Lett.} \textbf{\bibinfo{volume}{42}},
  \bibinfo{pages}{673} (\bibinfo{year}{1979}).

\bibitem{nandini_prl2004}
\bibinfo{author}{\bibfnamefont{D.}~\bibnamefont{Heidarian}} \bibnamefont{and}
  \bibinfo{author}{\bibfnamefont{N.}~\bibnamefont{Trivedi}},
  \bibinfo{journal}{Phys.\ Rev.\ Lett.} \textbf{\bibinfo{volume}{93}},
  \bibinfo{pages}{126401} (\bibinfo{year}{2004}).

\bibitem{scalettar_prb2007}
\bibinfo{author}{\bibfnamefont{P.~B.} \bibnamefont{Chakraborty}},
  \bibinfo{author}{\bibfnamefont{P.~J.~H.} \bibnamefont{Denteneer}},
  \bibnamefont{and} \bibinfo{author}{\bibfnamefont{R.~T.}
  \bibnamefont{Scalettar}}, \bibinfo{journal}{Phys.\ Rev.\ B}
  \textbf{\bibinfo{volume}{75}}, \bibinfo{pages}{125117}
  (\bibinfo{year}{2007}).

\bibitem{statDMFT_rwortis}
\bibinfo{author}{\bibfnamefont{Y.}~\bibnamefont{Song}},
  \bibinfo{author}{\bibfnamefont{R.}~\bibnamefont{Wortis}}, \bibnamefont{and}
  \bibinfo{author}{\bibfnamefont{W.~A.} \bibnamefont{Atkinson}},
  \bibinfo{journal}{Phys.\ Rev.\ B} \textbf{\bibinfo{volume}{77}},
  \bibinfo{pages}{054202} (\bibinfo{year}{2008}).

\bibitem{punnoose_sci2005}
\bibinfo{author}{\bibfnamefont{A.}~\bibnamefont{Punnoose}} \bibnamefont{and}
  \bibinfo{author}{\bibfnamefont{A.~M.} \bibnamefont{Finkel'stein}},
  \bibinfo{journal}{Science} \textbf{\bibinfo{volume}{310}},
  \bibinfo{pages}{289} (\bibinfo{year}{2005}).

\bibitem{imp_scaling_prb_2006}
\bibinfo{author}{\bibfnamefont{M.~C.~O.} \bibnamefont{Aguiar}},
  \bibinfo{author}{\bibfnamefont{V.}~\bibnamefont{Dobrosavljevi{\'c}}},
  \bibinfo{author}{\bibfnamefont{E.}~\bibnamefont{Abrahams}}, \bibnamefont{and}
  \bibinfo{author}{\bibfnamefont{G.}~\bibnamefont{Kotliar}},
  \bibinfo{journal}{Phys.\ Rev.\ B} \textbf{\bibinfo{volume}{73}},
  \bibinfo{pages}{115117} (\bibinfo{year}{2006}).

\bibitem{griffiths_2d}
\bibinfo{note}{E. C. Andrade, E. Miranda and V. Dobrosavljevi{\'c} (to be
  published)}.

\bibitem{nozieres74}
\bibinfo{author}{\bibfnamefont{P.}~\bibnamefont{Nozi\`eres}},
  \bibinfo{journal}{J. Low Temp. Phys.} \textbf{\bibinfo{volume}{17}},
  \bibinfo{pages}{31} (\bibinfo{year}{1974}).

\bibitem{milovanovic_prl1989}
\bibinfo{author}{\bibfnamefont{M.}~\bibnamefont{Milovanovi{\'c}}},
  \bibinfo{author}{\bibfnamefont{S.}~\bibnamefont{Sachdev}}, \bibnamefont{and}
  \bibinfo{author}{\bibfnamefont{R.~N.} \bibnamefont{Bhatt}},
  \bibinfo{journal}{Phys.\ Rev.\ Lett.} \textbf{\bibinfo{volume}{63}},
  \bibinfo{pages}{82} (\bibinfo{year}{1989}).

\bibitem{paalanen_prl1988}
\bibinfo{author}{\bibfnamefont{M.~A.} \bibnamefont{Paalanen}},
  \bibinfo{author}{\bibfnamefont{J.~E.} \bibnamefont{Graebner}},
  \bibinfo{author}{\bibfnamefont{R.~N.} \bibnamefont{Bhatt}}, \bibnamefont{and}
  \bibinfo{author}{\bibfnamefont{S.}~\bibnamefont{Sachdev}},
  \bibinfo{journal}{Phys.\ Rev.\ Lett.} \textbf{\bibinfo{volume}{61}},
  \bibinfo{pages}{597} (\bibinfo{year}{1988}).

\bibitem{vlad_kotliar_prl1993}
\bibinfo{author}{\bibfnamefont{V.}~\bibnamefont{Dobrosavljevi{\'c}}}
  \bibnamefont{and} \bibinfo{author}{\bibfnamefont{G.}~\bibnamefont{Kotliar}},
  \bibinfo{journal}{Phys.\ Rev.\ Lett.} \textbf{\bibinfo{volume}{71}},
  \bibinfo{pages}{3218} (\bibinfo{year}{1993}).

\bibitem{NFL_2005}
\bibinfo{author}{\bibfnamefont{E.}~\bibnamefont{Miranda}} \bibnamefont{and}
  \bibinfo{author}{\bibfnamefont{V.}~\bibnamefont{Dobrosavljevi{\'c}}},
  \bibinfo{journal}{Rep.\ Prog.\ Phys.} \textbf{\bibinfo{volume}{68}},
  \bibinfo{pages}{2337} (\bibinfo{year}{2005}).

\bibitem{tvojta_jpa}
\bibinfo{author}{\bibfnamefont{T.}~\bibnamefont{Vojta}}, \bibinfo{journal}{J.\
  Phys.\ A} \textbf{\bibinfo{volume}{39}}, \bibinfo{pages}{R143}
  (\bibinfo{year}{2006}).

\bibitem{seamus_davis_sci2005}
\bibinfo{author}{\bibnamefont{{{K. McElroy} {\it et al.}}}},
  \bibinfo{journal}{Science} \textbf{\bibinfo{volume}{309}},
  \bibinfo{pages}{1048} (\bibinfo{year}{2005}).

\bibitem{gargetal08}
\bibinfo{author}{\bibfnamefont{A.}~\bibnamefont{Garg}},
  \bibinfo{author}{\bibfnamefont{M.}~\bibnamefont{Randeria}}, \bibnamefont{and}
  \bibinfo{author}{\bibfnamefont{N.}~\bibnamefont{Trivedi}},
  \bibinfo{journal}{Nature Phys.} \textbf{\bibinfo{volume}{4}},
  \bibinfo{pages}{762} (\bibinfo{year}{2008}).

\end{thebibliography}

\end{document}